\documentstyle[12pt,a4,epsf]{article}
\newcommand{\D}{\partial}
\newcommand{\A}{\alpha}
\newcommand{\free}{\bar{\psi}i\D\!\!\!{/}\psi}
\newcommand{\F}{\frac}
\newcommand{\app}{\bar{\psi}}
\newcommand{\SR}{\!\!\!{/}}
\newcommand{\be}{\begin{equation}}
\newcommand{\ee}{\end{equation}}
\newcommand{\bea}{\begin{eqnarray}}
\newcommand{\eea}{\end{eqnarray}}
\newcommand{\DL}{\frac{\D}{\D\Lambda}}
\newcommand{\DT}{\frac{\D}{\D t}}

\newcommand{\DD}[3]{\frac{\overrightarrow{\delta}}{\delta #1}
#3 \frac{\overleftarrow{\delta}}{\delta #2}}
\newcommand{\DC}{\F{q^2}{C}\F{\D C}{\D q^2}}
\newcommand{\ST}{{\rm\bf Str}}

\newcommand{\Bold}[1]{#1}
\newcommand{\Bq}{{\bf q}}
\newcommand{\Bp}{{\bf p}}

\newcommand{\INTx}[1]{\int d^#1 x}

\newcommand{\FQ}{A_\mu,\app,\psi}

\newcommand{\TR}{^{\mbox{\tiny T}}}

\def\GV{G_{\mbox{\tiny V}}}

\def\JLone<#1,#2>{#1}
\def\JLtwo<#1,#2,#3>{#2}
\def\JLyear<#1,#2,#3,#4>{#3}
\def\JLpage<#1,#2,#3,#4>{#4}
\def\JL#1{\JLone<#1>\ {\bf \JLtwo<#1>} (\JLyear<#1>), \JLpage<#1>}
\def\Jpage<#1,#2,#3>{#3}
\def\andvol#1{{\bf \JLone<#1>} (\JLtwo<#1>), \Jpage<#1>}
\def\PTP#1{Prog.\ Theor.\ Phys.\ \andvol{#1}}

\def\PR#1{Phys.\ Rev.\ \andvol{#1}}

\def\PL#1{Phys.\ Lett.\ \andvol{#1}}
\def\NP#1{Nucl.\ Phys.\ \andvol{#1}}

\def\IJMP#1{Int.\ J.~Mod.\ Phys.\ \andvol{#1}}

\begin{document}
\begin{flushright}
KUCP0157\\
\end{flushright}
\begin{center}
{\large \bf
On the Running Gauge Coupling Constant\\
\vskip2mm
in the Exact Renormalization Group}
\end{center}
\centerline{Jun-Ichi Sumi}
\begin{center}
{\it
Department of Fundamental Sciences,
 Faculty of Integrated Human Studies,\\
 Kyoto University, Kyoto 606-8501, Japan}
\end{center}

\begin{abstract}
In this paper, we investigate the beta-function of the gauge coupling constant 
($e$) of the gauged four-fermi theory 
in the Exact Renormalization Group (ERG) framework. 
It seems that the presence of the four-fermi interaction strongly affects to 
the naive RG running of the gauge coupling constant. We show that this strong
correction has no physical meaning since the vertex $\app A\SR\psi$ 
involves the contribution from a mixing as well as a pure gauge interaction
due to the (anomalous) mixing among the photon and the vector composite field. 
By introducing the auxiliary field for the vector composite field, 
the situation turns to be rather clear. 
We adopt the counterterm to cancel the gauge non-invariant correction, 
and decompose the pure gauge interaction from the contribution of the mixing. 
We find the beta-function 
of the gauge coupling constant in the large $N$ limit.
\end{abstract}

\section{Introduction}
\label{sec:1}
The Exact Renormalization Group (ERG) \cite{rf:RGE,rf:FE}
is a powerful tool not only in the statistical physics but also in the 
particle physics i.e. the dynamical chiral symmetry breaking (D$\chi$SB) 
in the strong coupled gauge theory \cite{rf:chiral1,rf:chiral2,rf:chiral3}.
The ERG enable us to improve the ladder 
and/or the improved ladder (Higasigima) 
approximation \cite{SDmain}. In the ERG, one can easily 
incorporate the corrections from the non-ladder diagrams. 
In Ref.~\cite{rf:chiral2}, we have applied the ERG method 
to the chiral critical behavior in QED with the standing 
(constant) gauge coupling approximation. We have also seen that
the naive beta function of gauge coupling constant brings about 
the non-trivial ultra-violet stable fixed point. 
It shows the sharp contrast to the Gelmann-Low's RG beta-function 
of the gauge coupling constant which is positive semi-definite 
$\beta\ge0$\footnote{A Wilsonian RG $\beta$ function 
and a Gelmann-Low's one have opposite sign.} and has no ultra-violet (stable) 
fixed point \cite{rf:KUGO}. 
The additional fixed point in the ERG appears due to the breaking of the 
Ward-Takahasi identity, i.e. $Z_1=Z_2$. 
In the gauge invariant calculation, the running of the 
gauge coupling constant ($e$) is governed only by the photon's wave function 
renormalization, $Z_3$. However, in the ERG approach, 
the breaking of the Ward-Takahashi identity ($Z_1\ne Z_2$) also contributes 
to the beta-function of the gauge coupling $e$ and 
is enough large to change the qualitative feature of the continuum limit. 
Needless to say, we should carefully discuss this result. 

The D$\chi$SB in QCD was also investigated in Ref.~\cite{rf:chiral3}
partly by using the ERG. In those papers, however
the RG flow of the gauge coupling constant was that in 
the one-loop perturbation, not in the ERG. 
If one attempts to solve the D$\chi$SB 
only by the ERG, then one will encounter 
the problem due to the strong correction from the four-fermi
interaction, which is inconsistent with the gauge symmetry
to the RG beta-function of the gauge coupling constant. 
It is an obstacle to apply the ERG to solve D$\chi$SB. 

The ERG is the continuous version of 
the block spin transformation and one of the framework to perform the 
path-integrals. There are three formulation of the ERG, 
the Wegner-Houghton equation, the Polichinski equation and the evolution 
equation \cite{rf:RGE,rf:FE}. They are the functional differential equations 
for the Wilsonian effective action and/or the Legendre effective 
action with an infra-red cutoff $\Lambda$. 
The later is the one particle irreducible 
part of the Wilsonian effective action. 
In this paper, we employ the cutoff Legendre effective action 
$\Gamma_\Lambda[\Phi]$
\cite{rf:FE} since the Wilsonian effective action strongly depends on 
the cutoff scheme \cite{rf:MOR1,rf:scheme}. 

The infra-red cutoff is introduced as:
\be
S_{\rm cut}[A_\mu,\psi,\app]=\int d^4x\left(
\frac{\Lambda^2}{2}Z_3
A_\mu C^{-1}(-\partial^2/\Lambda^2) A_\mu +Z_2
\bar\psi C^{-1}_\psi(-\partial^2/\Lambda^2) i\partial\!\!\!/\psi\right),
\label{cutoffaction}
\ee
where $Z_2$ and $Z_3$ are the wave-function renormalization of 
the fermion and the photon respectively. 
The cutoff action $S_{\rm cut}$ preserves the chiral symmetry; 
$\psi\to e^{\theta\gamma_5}\psi$, and therefore $\Gamma_\Lambda[\Phi]$
also respect it. 
In this paper we do not specify the cutoff functions $C(x),C_\psi(x)$. 

As well known, a momentum cutoff which cannot be avoided to 
formulate the ERG, is not consistent with the gauge symmetry. 
Indeed, Eq.\ (\ref{cutoffaction}) conflicts with the gauge symmetry. 
Due to the renormalizability problem, the derivatives in $C_\psi$ 
cannot be replaced to the covariant ones $D_\mu=\D_\mu-ieA_\mu$. 
Thus to compensate the gauge invariance of the total solution, 
one has to introduce. The gauge non-invariant operators as the counterterms. 
theory space has to be enlarged to gauge non-invariant dimensions 
and next it should be restricted to the subspace maintaining 
the gauge symmetry of the total solution of the ERG. This process is tedious 
in general and demands more both human efforts and the computer 
resource. 

Recently the several attempts to construct the Wilsonian exact 
renormalization group consistent with the gauge symmetry are reported 
\cite{rf:gauge}. 
If one can construct it then the above problem is completely avoided. 
However, their formulations are not accompanied with the
non-perturbative approximation method and/or the recipe for extracting 
the physical information from `Wilsonian' effective action.
Hence we must chose either the gauge invariance or 
the non-perturbative approximation method/the above recipe. 
Hence for the practical and the non-perturbative analyses, 
it is necessary to solve the gauge non-invariant 
counterterms e.g. the photon mass term etc.. 

The effective action $\Gamma_\Lambda[\phi]$ satisfies the certain identity
at non-vanishing $\Lambda$ instead of the ordinary Slavnov-Taylor Identity
(STI), or the Ward-Takahashi identity. This identity is called 
the `Modified Slavnov-Taylor Identity' (MSTI) \cite{MSTI2}. 
The MSTI reduces to the STI in the infra-red limit; $\Lambda\to 0$
and ensures the gauge invariance of the total solutions of 
the ERG. The sub-space consistent with the MSTI can be regard
as the theory space of the gauge theory, i.e.
the Gauge Invariant Theory Space (GITS). 
It is in principle also possible to find the counterterms by the fine 
tuning the initial conditions to make the solution satisfy the STI. 
The result should coincide with the solution of the MSTI. 

\section{Exact renormalization group equation}
\label{sec:2}
Let us start from the ERG equation for
the cutoff Legendre effective action $\Gamma_\Lambda[\Phi]$,
\be
\Lambda\DL\Gamma_\Lambda[\Phi]=
\ST\left\{\left(\Bigl(\DC+1\Bigr)
{\bf 1}-\Bold{\eta}\right)\cdot
\left(1+{\bf C}\cdot\DD{\Phi\TR}{\Phi}{\Gamma_\Lambda[\Phi]}
\right)^{-1}\right\},
\label{eq:WET1}
\ee
where we use the condensed notation of the fields $\Phi\TR=(\FQ\TR)$ and of 
the anomalous dimensions $\Bold{\eta}={\rm diag}(\eta_A,\eta_\psi,\eta_\psi)$. 
One can easily generalize these to $N$ flavor case. 
The super trace $\ST$ involves both that of the Lorentz indices and the 
integral over the space-time coordinates. 
The matrix ${\bf C}$ is a following block diagonal matrix,
\be
{\bf C}^{-1}(\Bq)\equiv
\left(\begin{array}{c|c}
Z_3\cdot\Lambda^2\cdot C^{-1}\cdot\delta_{\mu\nu} & 
{\bf0} \\
      &      \\
\noalign{\vskip-3mm}
\hline
      &      \\
\noalign{\vskip-3mm}
{\bf 0} & \begin{array}{cc}
   0 & Z_2\cdot (\Lambda/q)^2C^{-1}\cdot q\SR \\
   Z_2\cdot (\Lambda/q)^2C^{-1}\cdot q\SR\TR  & 0
\end{array}
\end{array}
\right)\; ,
\ee
where $Z_3$ and $Z_2$ are the wave-function renormalization factors of the 
photon and of the fermion respectively. 
Now we choose the cutoff function $C(x)$ as 
the power like cutoff $C(x)=x^k$ : $k=1,2,\cdots$ \cite{rf:MOR2}. 
The anomalous dimensions above are given by, 
\be
2\eta_A=-\Lambda\DL \ln{Z_3},\qquad
2\eta_\psi=-\Lambda\DL \ln{Z_2}.
\ee

Next, we write all the dimensionful quantities in terms of the infra-red 
cutoff $\Lambda$, i.e. $\Phi=\Lambda^{d_\phi}\hat\Phi$, 
$\Bp=\Lambda\hat\Bp$ and ${\cal L}_\Lambda=\Lambda^d\hat{\cal L}_t$ 
where $d_\phi$, $t=\ln\Lambda_0/\Lambda$ and $\hat{\cal L}_t$ are the 
canonical dimension of the field; $d_\phi=(d-2)/2$, the 
cutoff scale factor and the action density respectively. We also 
write the dimensionless cutoff Legendre effective action 
$\hat \Gamma_t[\hat\phi]=\int d^d\hat x\hat{\cal L}_t(\hat\phi)$. Then, 
we have,
\be
\Lambda\F{\D}{\D\Lambda}\Gamma_\Lambda=
-\Lambda^d\left(\F{\D}{\D t}+d_\phi\Delta_\phi+\Delta_\D-d\right)
\hat \Gamma_t,
\label{eq:H}
\ee
where $\Delta_\phi$ and $\Delta_\D$ count the degree of the field and 
that of the derivative $\D_\mu$ respectively. 
Since the cutoff function ${\bf C}$ preserves the chiral symmetry, 
the effective action $\hat \Gamma_t[\hat\phi]$ also respects it. 

The initial boundary condition of the ERG flow equation is given by 
\be
\Gamma_{\Lambda=\infty}[\Phi]=S_{\rm bare}[\Phi],
\ee
and at $\Lambda=0$ the cutoff Legendre effective action
$\Gamma_\Lambda[\Phi]$ coincide with the ordinary effective action;
$\Gamma_{\Lambda=0}[\Phi]=\Gamma[\Phi]$.

\section{Structure of RG beta-functions in large N limit}
\label{sec:3}
Now let us consider $N$-flavor massless QED with the four-fermi operators. 
The fermionic field above is understood as $N$ Dirac fields, 
i.e. $\psi\longrightarrow\psi_i~~(i=1,\cdots,N)$. 
After rewriting the dimensionful parameters by the unit $\Lambda$, we write 
the initial effective action $\Gamma_0$ as,
\be
\Gamma_0[\Phi]=\INTx{d}\left\{
\frac{1}{4}Z_3 F_{\mu\nu}^2+\frac{1}{2\A}(\D_\mu A_\mu)^2
+\frac{1}{2}m^2 A_\mu^2+
Z_2\free+e \app A\SR\psi
-\frac{1}{2}\GV(\app\gamma_\mu\psi)^2\right\},
\label{effective}
\ee
where $m$ is the photon mass counterterm to cancel the gauge non-invariant 
correction to the photon mass. The gauge invariance requires the additional 
renormalization condition $m^2=0$ at $\Lambda=0$ or the Modified Slavnov-Taylor 
Identity (MSTI) \cite{MSTI2} at the finite cutoff $\Lambda\ne 0$. 

Let us consider the large $N$ limit, i.e.
\be
e^2\longrightarrow e^2/N,
\quad \GV\longrightarrow \GV/N, \quad N\longrightarrow\infty.
\ee
In this limit, RG flow of the four-fermi operator closes into the 
functional space $\{G^{\mu\nu}(P)\}$ as,
\be
-\frac{1}{2N}
\int \F{d^4P}{(2\pi)^4}(\app\gamma_\mu\psi)(P)
G^{\mu\nu}(P)
(\app\gamma_\nu\psi)(-P),
\ee
where $(\app\gamma_\mu\psi)(P)$ represents a Fourier transform of the 
local composite operator $\app(x)\gamma_\mu\psi(x)$. 
If we start from the action (\ref{effective}), 
the multi-fermi operators should take the form,
\be
\frac{1}{2N^{n-1}}
\int \prod_{i=1}^n\left(
\F{d^4P_i}{(2\pi)^4}(\app\gamma_{\mu_i}\psi)(P_i)\right)
G^{\mu_1\cdots\mu_n}(P_1,\cdots,P_{n-1})(2\pi)^4
\delta^4(\sum_i P_i),
\ee
and another multi-fermi operator like $\free(\app\gamma_\mu\psi)^2$
cannot appear. 
The RG beta-function for the momentum dependent four-fermi operator 
can be read,
\be
\DT G^{\mu\nu}(P)=-2G^{\mu\nu}(P)
-2P^2\F{\D}{\D P^2}G^{\mu\nu}(P)+G^{\mu\rho}(P)
I_{\rho\sigma}(P)G^{\sigma\nu}(P),
\ee
where $I_{\rho\sigma}(P)$ is a cutoff scheme dependent function 
given by, 
\be
I_{\mu\nu}(P)=
4(k+1-\eta_\psi)\int\F{d^4q}{(2\pi)^4}C(q)
S^2(q)C(P-q)S(P-q)
{\bf tr}[\gamma_\mu(P\SR-q\SR)\gamma_\nu q\SR],
\ee
and called threshold function. In the large $N$ limit we have $\eta_\psi=0$.
The function $S(q)$ corresponding to fermion's propagator is given by
\be
S(q)=1/(1+q^2C(q^2)).
\ee
We can decompose $G^{\mu\nu}(P)$ and $I_{\rho\sigma}(P)$ into two parts, 
the transverse part and the longitudinal one, i.e.
\bea
&&G^{\mu\nu}(P)=G_T(P)\left(g^{\mu\nu}-\F{P^\mu P^\nu}{P^2}
\right)+G_L(P)\F{P^\mu P^\nu}{P^2},\\
&&I_{\mu\nu}(P)=I_T(P)\left(g_{\mu\nu}-\F{P_\mu P_\nu}{P^2}
\right)+I_L(P)\F{P_\mu P_\nu}{P^2}
\eea
Then the RG flow equations of two parts of the four-fermi operator decouple
each other and we will find, 
\be
\DT G_{T,L}(P)=-2G_{T,L}(P)
-2P^2\F{\D}{\D P^2}G_{T,L}(P)+I_{T,L}(P)\cdot 
\left[G_{T,L}(P)\right]^2.
\label{eq:flow}
\ee
The higher operators do not contribute to the RG flow of $G_{T,L}(P)$ 
by a lack of the six-fermi operator in the large $N$ limit\footnote{%
The ERG flow equation involves at most the second functional derivative, 
no higher derivative with respect to the field. Therefore the
eight-fermi operator does not contribute to the RG flow of the four-fermi
operators.}. 

Next, let us consider the gauge sector. In large $N$ limit, 
$\app A\SR\psi$ interaction has the form\footnote{
By a virtue of the chiral symmetry, the Pauli term $F^{\mu\nu}\app
\sigma_{\mu\nu}\psi$ does not appear, therefore the
anomalous magnetic moment of the `electron' vanishes.},
\be
\int\F{d^4P}{(2\pi)^4}
A_\mu(P)\left[
\Gamma_T(P)\left(g^{\mu\nu}-\F{P^\mu P^\nu}{P^2}
\right)+
\Gamma_L(P)\F{P^\mu P^\nu}{P^2}
\right](\app\gamma_\nu\psi)(-P).
\ee
The photon two point function 
is also decomposed into two parts, the transverse part and the
longitudinal part,
\be
\F{1}{2}\int\F{d^4P}{(2\pi)^4}
A_\mu(P)\left[
\Pi_T(P)\left(g^{\mu\nu}-\F{P^\mu P^\nu}{P^2}
\right)+
\Pi_L(P)\F{P^\mu P^\nu}{P^2}
\right]A_\nu(-P).
\ee
The above functions $\Pi_{T,L}(P)$ and $\Gamma_{T,L}(P)$ have 
the following expansions,
\be
\left\{
\begin{array}{l}
\Pi_T(P)=m^2+Z_3P^2+\cdots,\\
\Pi_L(P)=m^2+\A^{-1}P^2+\cdots,\\
\Gamma_T(P)=e/\sqrt{N}+\cdots,\\
\Gamma_L(P)=e/\sqrt{N}+\cdots.
\end{array}
\right.
\ee
Here, the longitudinal parts $\Pi_L(P)$ and $\Gamma_L(P)-e_0/\sqrt{N}$ 
break the gauge symmetry, where $e_0$ is a bare gauge coupling. 
Hence, they should vanish at $\Lambda=0$. 
The quasi-locality of the threshold functions and of the bare action
\footnote{In another words, $I_{T,L}(P)$ are analytical at $P=0$.
} requires $\Pi_T(0)=\Pi_L(0)$ and $\Gamma_T(0)=\Gamma_L(0)$. 
Therefore the transverse part of the vacuum polarization at $P^2=0$
should vanish at $\Lambda=0$, i.e. $\Pi_T(P=0)=m^2=0$. 
The RG flow equations for $\Pi_{T,L}(P)$ and $\Gamma_{T,L}(P)$ 
can be found,
\bea
&&\DT \Pi_{T,L}(P)=\left(2
-2P^2\F{\D}{\D P^2}\right)\Pi_{T,L}(P)+I_{T,L}(P)\cdot
\left[\Gamma_{T,L}(P)\right]^2,
\label{eq:flow1}\\
&&\DT \Gamma_{T,L}(P)=
-2P^2\F{\D}{\D P^2}\Gamma_{T,L}(P)+I_{T,L}(P)\cdot 
\Gamma_{T,L}(P)\cdot G_{T,L}(P),
\label{eq:flow2}
\eea
where for convince, we deal with the vertices of the bare photon $A_\mu$
instead of that of the renormalized photon $\hat A_\mu=\sqrt{Z_3}A_\mu$. 

Now, let us consider the solutions of the RG flow equations (\ref{eq:flow}). 
It is more convenient to introduce the inverse of $G_{T,L}(P)$ i.e. 
$M_{T,L}(P)\equiv\left[G_{T,L}(P)\right]^{-1}$. Multiplying 
$-\left[M_{T,L}(P)\right]^2$ to both side of Eq.\ (\ref{eq:flow}), we find 
linear partial differential equations,
\be
\DT M_{T,L}(P)=2M_{T,L}(P)
-2P^2\F{\D}{\D P^2}M_{T,L}(P)-I_{T,L}(P).
\label{eq:flow3}
\ee
The initial boundary condition is given by $M_{T,L}(P)=1/\GV$ at $t=0$. 
One may easily find 
\be
\widetilde M_{T,L}(\widetilde P;\Lambda(t))=\F{1}{\GV}-
\int_0^tdt'{\rm e}^{-2t'}I_{T,L}({\rm e}^{2t'}\widetilde P^2),
\label{eq:largeNsol}
\ee
where $\widetilde M_{T,L}={\rm e}^{-2t}M_{T,L}$
and $\widetilde P={\rm e}^{-t}P$ are the inverse of the 
dimensionful four fermi vertex and the dimensionful momenta respectively. 
Since the last term of Eq.\ (\ref{eq:largeNsol}) corresponds to the 
fermionic bubble diagram, the inverse of 
$\widetilde M_{T,L}$ gives the famous chain sums. 

Especially for $\widetilde P=0$, we find,
\be
G_{T,L}^R=\GV/(1-I(P=0)\Lambda_0^2\GV).
\ee
If we take $\GV\to 1/I_{T,L}(P=0)\Lambda_0^2$, then $G_{T,L}^R$ 
blows up to infinity 
or equivalently $\widetilde M_{T,L}(\widetilde P=0;\Lambda=0)$ vanishs. 
Consequently, the four-fermi vertex acquires a massless pole in a 
vector channel\footnote{{\it Nota bene}, 
only a massless composite particle whose binding energy is zero can be
stable since the fermi fields remain massless due to the chiral symmetry.}.  
In the strong coupling region $\GV>1/I_{T,L}(P=0)\Lambda_0^2>0$, 
the true vacua breaks the Lorentz symmetry i.e. 
$<\app\gamma_\mu\psi>\ne0$, since $\widetilde M_{T,L}(0;\Lambda(t))$ 
correspond to the mass terms of the transverse and longitudinal modes 
of the vector composite field 
and they turn to a negative value in the strong coupling region. 

In our RG equation, when the vector 
four-fermi operator acquires a pole structure above, then 
$\Pi_{T,L}(P)$ should have a same pole structure due to 
the last term of Eq.\ (\ref{eq:flow2}). The longitudinal part of it 
conflicts to $Z_1=Z_2$, and should be compensated by a certain counterterm.
By the relations $\Pi_T(0)=\Pi_L(0)$ and $\Gamma_T(0)=\Gamma_L(0)$,
the constant part of transversal part $\Pi_T(P=0)$ should also vanish at 
$\Lambda=0$. 

\section{Solution in derivative expansion}
\label{sec:4}
In this section, we would like to explore the solutions in the 
derivative expansion. 
Let us restrict the sub-theory space to the following one. 
\bea
&&NG_T(P)=G_2 P^2+\GV,\qquad NG_L(P)=\GV,\\
&&\sqrt{N}\Gamma_T(P)=\Gamma_2 P^2+e,\qquad\sqrt{N}\Gamma_L(P)=e,\\
&&N^{-1}\Pi_T(P)=Z_3 P^2+m^2,\quad N^{-1}\Pi_T(P)=m^2,
\eea
where the $O(\D^0)$ parts of the longitudinal and the transversal vertices
should coincide due to quasi-locality of the threshold functions and 
the bare action. Inserting these and the truncated threshold functions; 
$I_T(P)=-P^2/6\pi^2+4\Omega,\;I_L(P)=4\Omega$, we have
\bea
&&\DT e=4\Omega e\GV,\label{lnbetae1}\\
&&\DT \GV=-2\GV+4\Omega \GV^2,\label{lnbetag1}\\
&&\DT G_2=-4G_2+8\Omega\GV G_2-\F1{6\pi^2}\GV^2,\label{lnbetag2}\\
&&\DT \Gamma_2=-2\Gamma_2+4\Omega
\left(eG_2+\Gamma_2\GV\right)-\F1{6\pi^2}e\GV,\label{lnbetae2}\\
&&\DT Z_3=\F1{6\pi^2}e^2-8\Omega e\Gamma_2.\label{wave2}
\eea
Note that, in the large $N$ limit the RG flow equations 
(\ref{lnbetae1})-(\ref{wave2}) forms a closed system in the total 
theory space, and therefore they describe exact results.\footnote{%
The approximation corresponding to that of Ref.\ \cite{rf:chiral2} 
will be found by setting $G_2=\Gamma_2=0$.}

The solution of Eqs.\ (\ref{lnbetae1})-(\ref{wave2}) can be found as 
follows. First, integrating Eq.\ (\ref{lnbetae1}), we have
\be
\GV=\F{\GV(0)}{2\Omega\GV(0)
+e^{2t}\left(1-2\Omega\GV(0)\right)},
\ee
where $\GV(0)$ is a bare four-fermi coupling constant. For another
coupling constants we have
\bea
&&e=\F{e_0}{1-2\Omega\GV(0)}(1-2\Omega\GV),\label{eq:a1}\\
&&G_2=-\F t{6\pi^2}\GV^2,\label{eq:a2}\\
&&\Gamma_2=\F{eG_2}{\GV},\label{eq:a3}\\
&&Z_3=1+\F{t}{6\pi^2}e^2,\label{eq:a4}
\eea
where we choose the initial boundary condition 
as; $G^{(2)}(0)=\Gamma^{(2)}(0)=0$, $e(0)=e_0$ and $Z_3(0)=1$.

Using these results, we can realized that the renormalized gauge
coupling constant $\hat e^2\equiv e^2/Z_3$ satisfies,
\be
\DT \hat e^2=-\F{\hat e^4}{6\pi^2}+8\Omega\left(
1-\F{t}{6\pi^2}\hat e^2\right)\hat e^2\GV.
\label{beta}
\ee
The `beta-function' of $\hat e^2$ 
has the positive region\footnote{%
Here the RG flow of the renormalized gauge coupling constant $\hat e$ 
depends on the cutoff scale parameter $t$ explicitly, since 
the shadow of the RG flow on the $\hat e-\GV$ plane does not draw 
the unique flow on the $\hat e-\GV$ plane
although the ERG flow on the full theory space; 
$\{\hat e,\GV,G_2,\Gamma_2\}$ does.}
 and the sign of the `beta-function' changes at
\be
\hat e^2=8\Omega(6\pi^2-\hat e^2 t)\GV.
\ee
The last term of Eq.\ (\ref{beta}) still breaks th WT identity and 
makes a dominant effect to the running of the `naive' gauge coupling
constant in both the ultra-violet and the infra-red regions.

\section{Introducing the Auxiliary field}
\label{sec:6}
Let us introduce the auxiliary field $V_\mu$ and give the counterterm for 
the mixing among $A_\mu$ and $V_\mu$ by the following Gaussian integral,
\be
{\cal N}=\int DV_\nu
\exp\left\{
-\int d^4x\frac{1}{2}M_{\rm V}^2\left(V_\mu
+\frac{1}{\sqrt{N}M_{\rm V}^2}\bar\psi_i\gamma_\mu\psi_i
+\theta A_\mu
\right)^2
\right\},
\ee
where $\theta$ is a certain constant determined by the gauge invariance. 
In the language of CJT effective action \cite{rf:CJT}, it is equivalent
to introduce the composite source $\Sigma_\mu$ for the vector
composite field as:
$\Sigma_\mu\cdot (
N^{-1/2}M_{\rm V}^{-2}\app_i\gamma_\mu\psi_i
+\theta A_\mu)$
except a physically non-important source quadratic term. 
The effective Lagrangian density ${\cal L}$ at the ultra-violet 
cutoff becomes,
\be
{\cal L}=\F{1}{4}\bar Z_3F_{\mu\nu}^2+\F{1}{2}M_A^2 A^2
+\F{\bar e}{\sqrt{N}}\bar\psi_iA\SR\psi_i
+\F{1}{2}M_{\rm V}^2 V^2+M_{\rm mixing}^2A\cdot V
+\F{1}{\sqrt{N}}\bar\psi_iV\!\SR\;\psi_i+\cdots,
\ee
where we introduce new variables, $\bar Z_3,M_A^2$ and $\bar e$ to 
distinguish from those in the previous section. Now we hold the ambiguity 
of the initial values of $M_{\rm V}^2$ and $y$ as $\GV=1/M_{\rm V}^2$
and $y=1$ at the ultra-violet cutoff $\Lambda_0$.

For the leading order in the derivative expansion, we have the RG equations;
\bea
&&\DT \bar e=\DT y=0,\\
&&\DT
 M_A^2=2M_A^2-4\bar e^2\Omega,\\
&&\DT
 M_{\rm V}^2=2M_{\rm V}^2-4\Omega,\\
&&\DT
 M_{\rm mixing}^2=2M_{\rm mixing}^2-4\bar e\Omega.
\eea
Here, the RG flow equations of $O(\D^0)$ coupling constants are not 
affected by the higher derivative couplings. 
By the condition $\GV=1/M_{\rm V}^2$, the four-fermi coupling is not 
generated, since once $\GV$ vanishs then the 
beta-function of $\GV$ is also vanish for each scale\footnote{%
The RG flow of $\GV$ is also give by Eq. (\ref{lnbetag1}).}.  
Hence at all scale we have $\GV=0$ and  
the beta-function of $e$ does not have the anomalous region. 
Therefore the physical running of the gauge coupling 
is governed only by the wave-function renormalization of the gauge field. 

To see the RG running of the physical gauge coupling, we must 
calculate the wave-function renormalizations. 
Let us write the kinetic terms of the gauge field and composite 
vector field as:
\be
{\cal L}\sim
\F{1}{4}\bar Z_3F_{\mu\nu}F_{\mu\nu}+
\F{1}{2}Z_MF_{\mu\nu}G_{\mu\nu}+
\F{1}{4}Z_VG_{\mu\nu}G_{\mu\nu}+\cdots,
\ee
where $G_{\mu\nu}$ is a field strength of the vector composite field
$V_\mu$, i.e. $G_{\mu\nu}=\D_\mu V_\nu-\D_\nu V_\mu$. 
The gauge invariant kinetic term mixing $Z_M$ does not affect the 
RG flow of the gauge coupling since the rotation diagonalizing 
the kinetic terms, should take a form by the gauge transformation low of
the gauge field, 
\be
\left(\begin{array}{c}
A_\mu\\
V_\mu
\end{array}\right)=
\left(\begin{array}{cc}
1&\delta_1\\
0&\delta_2
\end{array}\right)
\left(\begin{array}{c}
\widetilde A_\mu\\
\widetilde V_\mu
\end{array}\right),
\ee
which preserves the $Z$-factor of the gauge field. We find the RG
equation for $Z_3$ as,
\be
\DT\bar Z_3=\F{\bar e^2}{6\pi^2},
\ee
and the solution $\bar Z_3=t\bar e^2/6\pi^2$.

Next, the counterterm $M_{\rm mixing}^2(\bar e)$ should be constrained 
by
\be
\left(\DT\bar e\right)
\F{\partial}{\partial \bar e}M_{\rm mixing}^2(\bar e)
=\DT M_{\rm mixing}^2,
\label{eq:mix}
\ee
with a boundary condition $M_{\rm mixing}^2\to 0$ as $e_R\equiv
\bar e/\sqrt{\bar Z_3}\to 0$\footnote{It is similar to the coupling
reduction \cite{rf:couplinreduction}.}. One can easily get 
$M_{\rm mixing}^2=2\bar e\Omega$. In the same manner, we also 
find $\widetilde M_A^2=2\bar e^2\Omega$ for the photon mass counterterm. 

The relations between the original coupling constants i.e. before introducing 
the auxiliary field $\Gamma_t[A_\mu,\app,\psi]$ ant those after introducing
the auxiliary field $\Gamma_t[A_\mu,\app,\psi,V_\mu]$ can be easily found.
First for the gauge coupling constant $e$ we have,
\be
e=\bar e-\left[
\F{M_{\rm mixing}^2}
{M_{\rm V}^2}\right]=\bar e(1-2\Omega \GV). 
\label{eq:map}
\ee
The mappings among another parameters e.g. the wave-function renormalizations 
also can be found. For each scale $t$, the auxiliary field $V_\mu$ can 
be integrated out since the loop corrections of $V_\mu$ are dropped in 
the large $N$ limit. Taking account of the tree diagrams we have the 
relations;
\bea
&&Z_3=\bar{Z_3}+\left[\F{M_{\rm mixing}^4}{M_{\rm V}^4}Z_V
-2\F{M_{\rm mixing}^2}{M_{\rm V}^2}Z_M\right],\label{eq:r1}\\
&&m^2=\bar{M_A^2}-\left[\F{M_{\rm mixing}^4}{M_{\rm V}^2}\right],\\
&&G_2=-\F{Z_V}{M_{\rm V}^4},\label{eq:r2}\\
&&\Gamma_2=\left[\F{M_{\rm mixing}^2}{M_{\rm V}^4}
Z_V\right]-\F{Z_M}{M_{\rm V}^2}.\label{eq:r3}
\eea
One can find the solutions (\ref{eq:a1})-(\ref{eq:a4}) by using
$\bar Z_3/\bar e^2=Z_M/\bar e=Z_V=t/6\pi^2$, $\GV=1/M_V^2$, 
$M_{\rm mixing}^2=2\bar e\Omega$ and $e_0=\bar e(1-2\Omega\GV(0))$.
The terms in the square bracket are the contributions from the gauge 
non-invariant corrections, since these proposal to $M_{\rm mixing}^2$. 
In the language of the ordinary perturbation theory, $M_{\rm mixing}^2$
corresponds to a quadratically divergent one-loop contribution 
like a photon's mass correction. As well-known, such a correction 
breaks the WT identity. The gauge invariant vertices also suffer from
the such correction through quadratically diverging renormalization parts, i.e.
for example, the terms in the square bracket in Eqs.\ (\ref{eq:r1}),
(\ref{eq:r3}) and (\ref{eq:r3}).
The anomalous running of $e$ in Eq.\ (\ref{beta}) 
is essentially a result form these corrections. By introducing the
auxiliary field $V_\mu$, we can decompose these contributions 
from the gauge invariant contributions i.e. $\bar e,\bar{Z_3}$ etc..

Hence the strong correction of the RG flow of the gauge coupling constant 
from the $e\GV$ term is physically meaningless. 
It is due to the fact that the 
$A_\mu\app\gamma_\mu\psi$ vertex is not a purely gauge interaction but 
including a mixing among the gauge field and the vector composite 
field and that the wave-function renormalization factor $Z_3$ also 
suffers from a mixing. 

In the large $N$ limit, we introduced the auxiliary field for 
the vector composite field $V_\mu\sim\app\gamma_\mu\psi$.
Then we could easily distinguish the pure gauge interaction 
from the mixing among the gauge field and the vector composite 
field. After resolving a mixing, we found the correct RG running of the 
gauge coupling constant $e$. 

The MSTI leads to the relation corresponding to Eq.\ (\ref{eq:map})
but another relations, for example Eq.\ (\ref{eq:r1}), since they
are the gauge invariant vertices. The MSTI tells the informations for
the counterterms compensating the gauge invariance but the recipes to
distinguish the contribution from the gauge non-invariant sub-diagrams
from the purely gauge invariant corrections.

For a finite $N$, we cannot easily solve a mixing therefore 
the analyses will be more complicated. An essential difference from 
the case of the large $N$ limit is the propagation of the auxiliary fields.
The $e\GV$ term is composed of a mixing 
among the vector composite field and a propagation of the vector (scalar) 
composite field $\app\gamma_\mu\psi$ ($\app\psi$). 
The later one should also contribute to 
the wave-function renormalization factor ($Z_2$) of the fermion, 
and cancel to each other. To see this, we must introduce an 
auxiliary field for the scalar composite field too. 

For the non-abelian case like QCD, there are no physical particle 
which has a same quantum number with gluons since the colored particles
should be confined. Hence the vertex $A_\mu^a\app T^a\gamma_\mu\psi$ 
does not have a pole structure in a strict sense. 
However the gauge non-invariant renormalization parts will 
affect the Wilsonian RG flow of the gauge coupling constant. 

\section*{Acknowledgements}

The author thanks to K-I. Aoki, H. Terao, K. Morikawa, W. Souma
 and M. Tomoyose for the valuable discussions and comments.

\end{document}